\documentstyle[11pt,newpasp,twoside,epsf]{article}
\def\edcomment#1{\iffalse\marginpar{\raggedright\sl#1\/}\else\relax\fi}

\reversemarginpar

\begin{document}
\title{Prospecting for Heavy Elements with Future Far-IR/Submillimeter Observatories}
 \author{D. Leisawitz}
\affil{NASA GSFC, Code 631, Greenbelt, MD 20771}
\author{D.J. Benford}
\affil{NASA GSFC, Code 685, Greenbelt, MD 20771}
\author{A. Kashlinsky}
\affil{SSAI and NASA GSFC, Code 685, Greenbelt, MD 20771}
\author{C.R. Lawrence}
\affil{JPL, MS 169-327, 4800 Oak Grove Dr., Pasadena, CA 91109}
\author{J.C. Mather}
\affil{NASA GSFC, Code 685, Greenbelt, MD 20771}
\author{S.H. Moseley}
\affil{NASA GSFC, Code 685, Greenbelt, MD 20771}
\author{S.A. Rinehart}
\affil{NRC RRA, NASA GSFC, Code 685, Greenbelt, MD 20771}
\author{R.F. Silverberg}
\affil{NASA GSFC, Code 685, Greenbelt, MD 20771}
\author{H.W. Yorke}
\affil{JPL, MS 169-506, 4800 Oak Grove Dr., Pasadena, CA 91109}

\begin{abstract}
To understand the cosmic history of element synthesis it will be important 
to obtain extinction-free measures of the heavy element contents of high-redshift objects and to chart two monumental events: the collapse of the first metal-free 
clouds to form stars, and the initial seeding of the universe with dust. The information needed to achieve these objectives is uniquely available in the far-infrared/submillimeter (FIR/SMM) spectral region. Following the Decadal Report 
and anticipating the development of the Single Aperture Far-IR (SAFIR) telescope 
and FIR/SMM interferometry, we estimate the measurement capabilities of a large-aperture, background-limited FIR/SMM observatory and an interferometer 
on a boom, and discuss how such instruments could be used to measure the element synthesis history of the universe.
\end{abstract}

\section{Importance of the FIR/SMM Spectral Region}

Prior to enrichment with heavy elements the first collapsing protostellar clouds must have given up the bulk of their potential energy through the most readily excited H$_{2}$ lines, namely those at the rest wavelengths 17 and 28 \micron\ (Haiman, Rees, \& Loeb 1996).  Heavy elements forged in the first stars polluted the intergalactic and interstellar medium and were incorporated into later stellar generations. Indeed, these elements must have had an effect on the star formation process through their strong influence on the energy balance of the interstellar medium. Among the most prominent manifestations of heavy elements in the rest frame spectrum of a galaxy are the far-IR C, N, and O atomic and ionic fine structure lines, the CO lines from molecular clouds, the far-IR thermal continuum emission from interstellar dust, and the mid-IR [Ne II] and [O IV] lines and PAH features from gas and dust. 

The [C II] line at 158 \micron , the [O I] lines at 63 and 146 \micron , the [OIII] lines at 52 and 88 \micron , and the [N II] lines at 122 and 205 \micron\ have potential to serve as metallicity tracers out to very high redshifts, sampling gas exposed to various intensities of ionizing radiation. The 158 \micron\ line alone contains $\sim 0.1$ -- 1\% of a galaxy's bolometric luminosity (Stacey et al. 1991). It is the strongest line in the spectrum of our own galaxy, and the nitrogen lines are also quite strong (Wright et al. 1991; Bennett et al. 1994). Oxygen is the most abundant heavy element, and emission can be expected to be seen in the atomic or ionic O lines depending on the nature of the source. For example, the [O III] lines are bright in IR-luminous galaxies (Fischer et al. 1999). Although the relative intensity of the 158 \micron\ line varies from galaxy to galaxy (Fischer et al. 1999), independent diagnostics of the excitation conditions will aid in the interpretation of the line strength, and a good physical model exists for the emission (Wolfire et al. 1989). Significantly, the far-IR lines do not suffer the ill effects associated with extinction, whereas large and uncertain extinction corrections are required at UV/optical wavelengths. Several mid-IR fine-structure lines have been shown to be good diagnostics of the radiation field hardness and emission source (Lutz et al. 1999; Thornley et al. 2000), and the [O III] line ratio can be used to measure the electron density in the emitting region (Fischer et al. 1999). Thus, multi-line studies in the FIR/SMM can be used reliably to trace metallicity as a function of redshift out to $z \sim 5$, and these lines can be followed out to even greater redshifts at millimeter wavelengths with ALMA.
 
Approximately half of a typical spiral galaxy's luminosity is emitted in the far-IR because interstellar dust absorbs UV and visible starlight and radiates thermally at these longer wavelengths. The redshifted dust continuum emission peak would remain in the FIR/SMM spectral range for galaxies out to $z \sim 10$ if galaxies exist there and contain dust. Evidence for dust has been seen in a galaxy at $z = 5.34$ (Armus et al. 1998), and we would like to know when the first dust formed.

Ultraluminous IR galaxies (ULIRGs) exhibit strong features in their mid-IR spectra, likening them in this respect to starburst galaxies and not AGN (Lutz et al. 1999). While their possible evolutionary connection to quasars makes ULIRGs intrinsically interesting, for the sake of this discussion they warrant attention simply because they are luminous and the spectral features are commonly attributed to emissions from polycyclic aromatic hydrocarbons (PAHs). The 7.7 \micron\ PAH feature has good potential as a tracer of heavy elements out to great distances.

\section{Future FIR/SMM Space Observatories and Their Measurement Capabilities}

Future FIR/SMM space observatories will have the sensitivity needed to detect high-$z$ objects, the angular resolution needed to separate the emissions of individual galaxies, and the spectral resolution needed to measure line luminosities and calculate heavy element abundances. In this section we give estimates of the sensitivity of SAFIR and a FIR/SMM interferometer to line emission, PAH spectral features, and the thermal dust continuum emission from high-redshift galaxies.

SAFIR is a filled-aperture far-IR telescope (Rieke et al. 2002) recommended for an end-of-the-decade start in the Decadal Report. As conceived, SAFIR would be background limited over a wavelength range from about 15 to 600 \micron\ to overlap slightly with NGST and ground-based capabilities, and would be diffraction limited at $\sim 40$ \micron . With a 10 m aperture, and with new generations of cold detectors, it would have 150 times the collecting area and an order of magnitude greater angular resolution at a given wavelength than SIRTF. SAFIR instruments would provide imaging and spectroscopic capabilities with maximum spectral resolution $R = \lambda / \Delta \lambda \sim 10^6$, much greater than that needed for the extragalactic studies discussed here. To achieve the goal of background-limited performance the SAFIR mirror will have to be cooled to about 4 K and the detector NEP will have to be limited to $\sim 10^{-20}$ W Hz$^{-1/2}$. 

The Decadal Report also recommends the development of enabling technologies for FIR/SMM interferometry, recognizing this as a compelling and promising future direction (Leisawitz et al. 2002). In the present context an interferometer is important because it could make spectral line measurements complementary to those of SAFIR at the longer FIR/SMM wavelengths, where SAFIR will be confusion noise limited. Because this paper emphasizes spectral measurements of distant galaxies rather than imaging, we don't say much about the very long baseline interferometer known as SPECS (the Submillimeter Probe of the Evolution of Cosmic Structure) and instead discuss the Space Infrared Interferometric Telescope (SPIRIT), a FIR/SMM spatial and spectral (i.e., ``double Fourier''; Mariotti \& Ridgway 1988) interferometer on a boom, and a likely forerunner to SPECS. SPIRIT could have a maximum baseline length in the 30 -– 50 m range and provide sub-arcsecond angular resolution out to sub-millimeter wavelengths. Like SPECS, SPIRIT is conceived as a background-limited Michelson stellar interferometer with a scanning optical delay line (Mather et al. 2002). The delay line stroke length can be set to provide R = 1000 spectroscopy, and the light collecting mirrors can be moved to provide dense {\it u-v} plane coverage, and thus imaging capability.

How sensitive can we expect the future FIR/SMM space observatories to be? Both SAFIR and SPIRIT are supposed to be background limited. We use an empirical model based on COBE observations (Hauser et al. 1998) to calculate the FIR/SMM diffuse background, and Blain's (2000) model for the confusion noise due to extragalactic sources and Galactic cirrus clouds (Gautier et al. 1992). The diffuse emission model includes zodiacal, Galactic cirrus, and CMB components. 

To estimate the sensitivity of SAFIR we assume that the telescope is equipped 
with an R = 1000 spectrometer, high QE detectors ($\eta = 0.5$), and has total system transmission $\tau_{sys} = 0.3$. The minimum detectable line flux ($1 \sigma$) for a background-limited telescope of aperture area A in exposure time t is $MDLF = NEP_{bg} (A \tau_{sys} t^{1/2})^{-1}$, where the noise equivalent power attributable to background photons that reach the detector, 
$NEP_{bg} = (2 P_{bg} h \nu / \eta)^{1/2}$, and $P_{bg} = (\nu I_{\nu})_{bg} A \Omega \tau_{sys} / R$. The etendue $A \Omega \sim \lambda^2$ at wavelength $\lambda$. The confusion noise for a 10 m diffraction limited telescope in a spectral channel $\Delta \nu = \nu / R$ is added in quadrature with this background-limited MDLF to estimate the line flux sensitivity shown in Figure 1. Confusion is expected to be the dominant noise source in any reasonable exposure time at $\lambda \ga 200$ \micron . In this paper we ignore the possibility that line emission below the confusion limit might be attributable to a particular source if its redshift is independently determined. The effect of confusion is illustrated in Figure 2.

\begin{figure}
\plotone{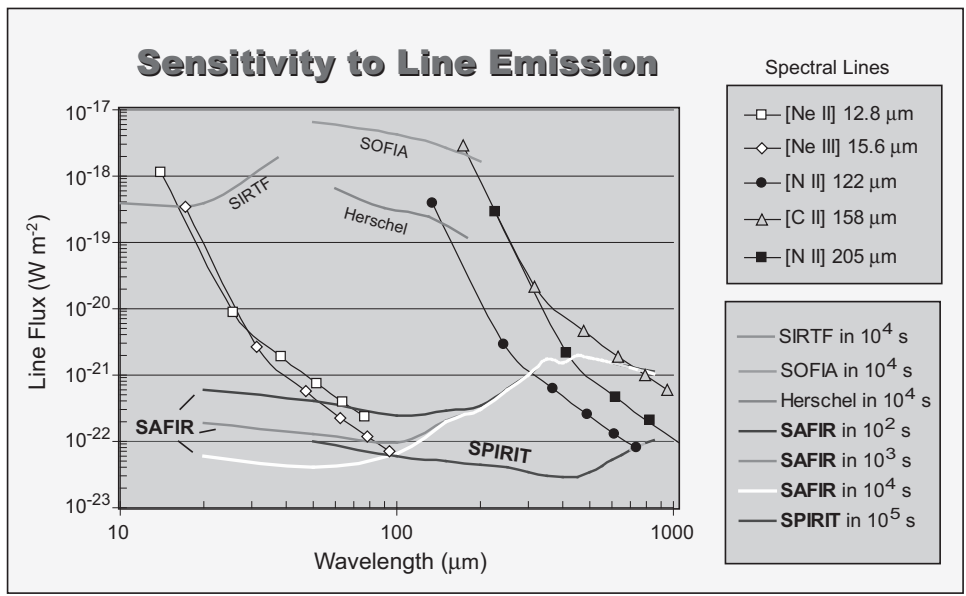}
\caption{A spectrometer on SAFIR (with assumed $\lambda / \Delta \lambda = 10^3$) 
would be 4 to 5 orders of magnitude more sensitive than the corresponding instruments on SIRTF, SOFIA, and Herschel, enabling unprecedented studies of the star formation process and astrophysical conditions in distant, young galaxies. Estimated strengths of five important diagnostic and interstellar gas cooling lines are shown for a hypothetical ``Milky Way'' galaxy ($L_{FIR} = 2 \times 10^{10} L_{\sun}$) at redshifts of 0.1, 1, 2, 3, 4, and 5 (symbols along each curve, with redshift increasing from the upper left to the lower right). The rest wavelengths of the spectral lines are given in the inset. SAFIR could, for example, measure the [Ne II] and [Ne III] lines in ``normal'' galaxies out to $z = 5$ in modest exposure times. The relative intensity of these lines can be used to discriminate between AGN-dominated and star formation-dominated emission. Many galaxies are much more luminous than the Milky Way, making them even easier to see. At $\lambda > 200$ \micron , SAFIR would reach the confusion noise ``floor'' in about 100 seconds; longer exposure times would not help. However, because of its greater resolving power and still substantial total aperture area, the SPIRIT interferometer would break the confusion barrier and probe the universe to comparable depth (redshift $z \sim 5$) in the spectral lines that dominate the cooling of interstellar gas and allow the gas clouds to collapse and produce stars.}
\end{figure}

\begin{figure}
\plotone{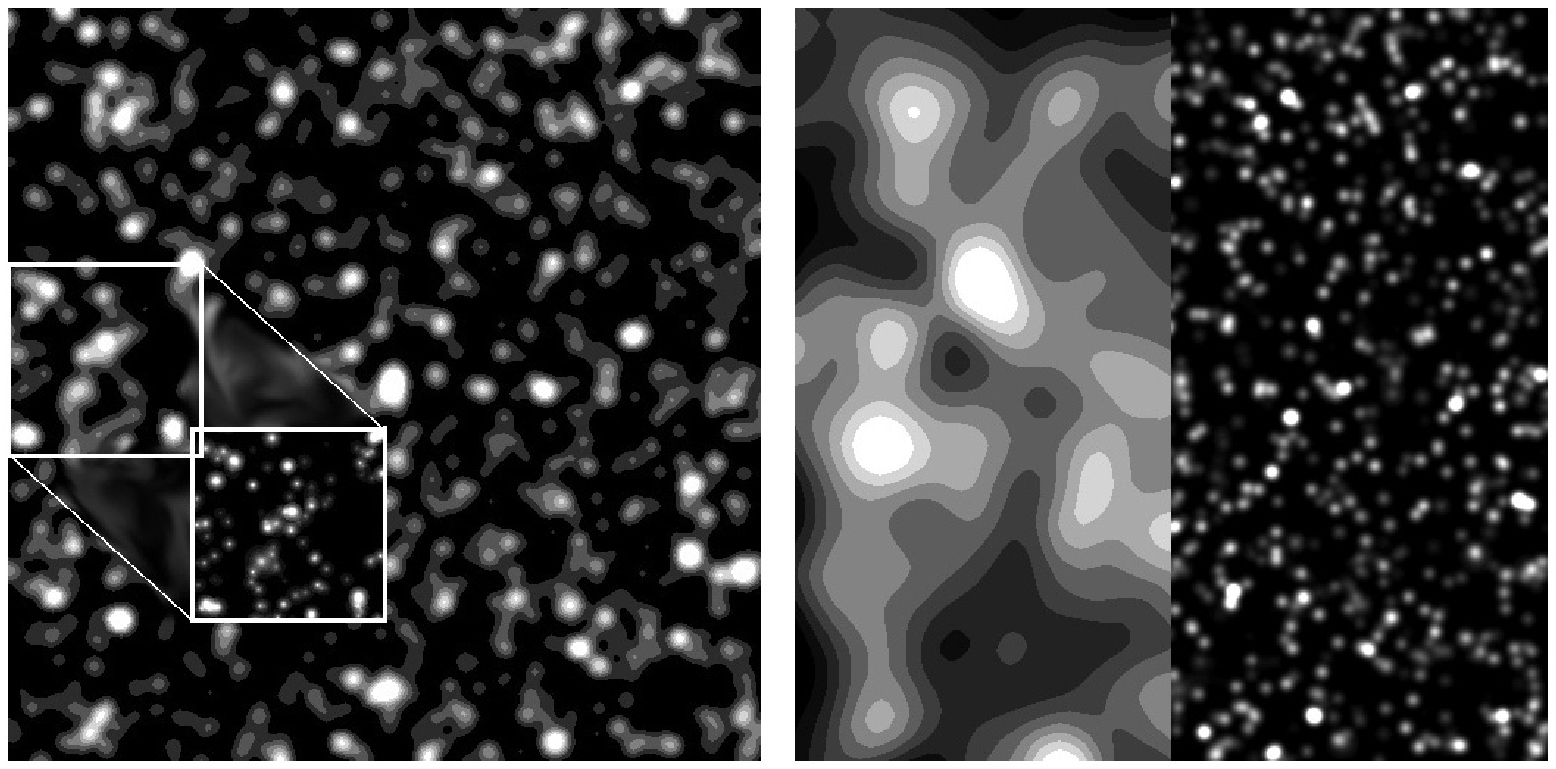}
\caption{Simulated 100 \micron\ (left) and 450 \micron\ (right) images of a 2.4 arcmin square high Galactic latitude region. The simulation includes zodiacal and Galactic foreground emission components and extragalactic sources distributed in luminosity and redshift according to Blain's (2000) model. The 100 \micron\ image shows the sky at SAFIR resolution, and the lower-right inset shows a small part of the field at full simulated resolution (prior to convolution with the SAFIR beam). The same field is shown in the 450 \micron\ image at SAFIR resolution on the left and SPIRIT resolution on the right. SAFIR is only modestly affected by confusion at 100 \micron\ but severely affected at 450 \micron . SPIRIT beats confusion at 450 \micron .}
\end{figure}

To a good approximation, the MDLF for SPIRIT is related to the background-limited 
MDLF given above for SAFIR by the factor $(A_{SAFIR} / A_{SPIRIT}) n^{1/2}$, in which $A_{SAFIR} = (\pi / 4) (10 m)^2$, $A_{SPIRIT}$ is the total light collecting aperture of SPIRIT, and $n$ is the number of collecting mirrors. The factor $n^{1/2}$ allows for the fact that independent, uncorrelated samples of background photons reach the individual collecting mirrors. To derive the sensitivity curve for SPIRIT shown in Figure 1 we assume that the interferometer has two 3 m diameter light collecting mirrors. SPIRIT would not be confusion limited at any wavelength in $10^5$ s.

Figure 1 shows that a line flux of $\sim 10^{-22}$ W m$^{-2}$ is detectable with SAFIR or SPIRIT in a reasonable amount of time anywhere in the FIR/SMM spectral range. Due to its cold temperature and large aperture size SAFIR would be 4 to 5 orders of magnitude more sensitive to line emission than SIRTF, SOFIA, and Herschel. Despite its smaller total aperture area than SAFIR, SPIRIT would beat confusion (Fig. 2) and have comparable sensitivity. The Atacama Large Millimeter Array (ALMA) will provide a similar capability at millimeter wavelengths. From a still broader perspective, nearly all the light emitted by luminous objects from the present back in time to the epoch of their first formation will be accessible to future astrophysicists with NGST, SAFIR, SPIRIT, and ALMA.

Figure 1 also shows the predicted sensitivity of SAFIR and SPIRIT to prominent C, N, and Ne lines emitted by normal galaxies at various redshifts. The mid-IR Ne diagnostic lines could be followed beyond the NGST spectral range with SAFIR out to $z > 5$, yielding a measure of the excitation conditions, and thereby aiding in the interpretation of C, N, and O line observations. The [C II] 158 \micron\ line would be easily detected with SAFIR out to $z = 3$, with SPIRIT out to $z = 4$, and with ALMA at least out to $z = 5$. Confusion noise might prevent SAFIR from observing the [N II] lines beyond $z \sim 1.5$ or 2, and likewise the oxygen lines mentioned previously, but SPIRIT could measure most of these lines all the way to $z = 5$, and ALMA would pick up the 205 \micron\ [N II] line when it exits the SPIRIT spectral range at $z \sim 3$. All of these estimates of detectability pertain to a galaxy of relatively modest luminosity; the Milky Way spectrum (Wright et al. 1991; Bennett et al. 1994) was used as the standard. An IR-luminous galaxy can be two or more orders of magnitude more luminous than the Milky Way, which would bring the [O III] 52 and 88 \micron\ lines, for example, well within the SAFIR sensitivity range for galaxies out to $z = 5$. 

There is a reasonable chance that SAFIR could unveil the H$_2$ line emission from the clouds that collapsed to form the very first generation of stars. The likely presence of a pair of lines at rest wavelengths 17 and 28 \micron\ would compensate for the fact that the epoch of initial star formation, and therefore the redshift, is unknown. Recent estimates place the emission in the redshift range $z \sim 10 - 20$ (Loeb \& Barkana 2001). Thus, fortuitously, the lines would appear in a spectral range where the background noise is nearly at a minimum, enhancing the prospects for detection. The H$_2$ line strength can be estimated as follows. If the protostellar clouds begin at $T \sim 10^3$ K and $n_{H} = 100 \, {\rm cm}^{-3}$ they would cool on a timescale of a few Myr. The corresponding cloud luminosity is only about 0.01 L$_{\sun}$, but many such clouds could be concentrated in an object as massive as a small galaxy ($10^{10}$ M$_{\sun}$), and all of this flux would be concentrated in a single SAFIR beam. We ignore Thompson scattering by electrons freed up in the reionization of the universe because the scattering optical depth is likely $< 0.2$ (D. Spergel, private communication). If all the collapsing cloud energy is apportioned about equally between 17 and 28 \micron\ lines, then the line strengths would be in the $10^{-23}$ to $10^{-24}$ W m$^{-2}$ range. If the epoch of initial cloud collapse was short-lived, then the lines might be found serendiptously when many spectra obtained for other purposes are coadded. If the lines are lost in confusion noise to SAFIR, then there is a chance that SPIRIT or the longer baseline interferometer SPECS would see them. The detection of this emission would be a major coup, as it would tell us when the clock began to tick for heavy element formation and reionization.

The sensitivity of SAFIR to continuum emission or resolved spectral features is given by the "minimum detectable flux density," $MDFD = MDLF / \Delta \nu$, where $\Delta \nu$ could be $\sim \nu / 5$ if one were interested in the continuum spectral energy distribution (SED), or $\sim \nu / 100$ if one were interested in resolving, say, the 7.7 \micron\ PAH feature. Although the sensitivity of an interferometer of the type described above would be further compromised by efficiency factors if its full spectral resolution capability is not exploited, for SPIRIT the loss would be small, and we can scale by $(A_{SAFIR} / A_{SPIRIT}) n^{1/2} \simeq 8$, as before, to estimate the MDFD.  

The dust continuum emission from a normal galaxy ($L_{FIR} = 2 \times 10^{10}$ L$_{\sun}$) at $z = 2$ would peak at about 50 $\mu$Jy at 250 \micron\ in the observer's reference frame. SAFIR could detect this signal in a very short exposure time, but it would be right at the confusion limit. Although SAFIR would have ample sensitivity to detect continuum emission from normal galaxies well into the Wien regime, where, for a given redshift, the wavelength is shorter and the angular resolution of the telescope is better, the bulk of the emission would become shrouded by confusion when the galaxy is at $z \ga 2$. Better resolution is the key to going deeper. SPIRIT could easily measure the dust continuum SEDs of normal galaxies out to $z = 5$. 

IR-luminous galaxies are another story. Such galaxies may have been common in the early universe, and they tend to be warmer than normal galaxies. Lutz et al. (1999) combine ISO measurements of 60 ULIRGs, normalize the spectra to $S_{\nu}(60 \micron ) = 1$ Jy, and find that the average spectrum contains a PAH 7.7 \micron\ feature with peak flux density 0.01 Jy. The PAH and 60 \micron\ continuum flux density from a $10^{12}$ L$_{\sun}$ ULIRG, the least powerful source of its class, would be easily detectable with SAFIR out to $z > 5$. By measuring these emissions from ULIRGs at many different redshifts one might find a trend that reveals the buildup of dust over time. SIRTF will add considerably to our understanding of PAH production and aid in our interpretation of the mid-IR spectral features.

\section{Summary and Conclusions}

Information vital to our understanding of the cosmic history of element synthesis is uniquely available at far-infrared/submillimeter wavelengths. To extract this information we will need telescopes and instruments that provide spectral line sensitivity of the order of $10^{-22}$ W m$^{-2}$ at spectral resolution $R \ga 1000$, and $\mu$Jy sensitivity to continuum emission, over the spectral range 15 -- 800 \micron . Instruments on SIRTF, SOFIA, and Herschel will operate in this range, but with orders of magnitude less sensitivity than is required to measure the dominant cooling and important diagnostic spectral lines, the dust continuum emission, and the PAH features in the spectra of young galaxies and protogalactic objects at high redshifts. Measurement capabilities comparable to those that will become available at shorter wavelengths with NGST and at longer wavelengths with ALMA will be needed in the FIR/SMM. It is envisioned in the Decadal Report that these capabilities will be provided with the Single Aperture Far-IR (SAFIR) telescope and a far-IR interferometer. 

To probe the heavy element and dust formation history of the universe SAFIR should carry instruments that provide medium resolution ($R \sim 10^3$) spectroscopy over the wavelength range 15 -- 600 \micron , and low-resolution spectroscopy ($R \sim 10^2$) over the 15 -- 50 \micron\ spectral range. SAFIR should be about 10 m in diameter, not only to provide adequate sensitivity at high spectral resolution (for purposes other than those considered here), but also to provide a view of the extragalactic IR sky unlimited by confusion at wavelengths $< 200 \micron $.

At longer wavelengths, confusion noise will limit the sensitivity of SAFIR, so the Space Infrared Interferometric Telescope (SPIRIT) should be designed to break the FIR/SMM confusion barrier and make complementary spectroscopic and spectrophotometric measurements of high-$z$ objects. An interferometer with a maximum baseline length in the 30 -- 50 m range would beat confusion. Due to its smaller total aperture area and mode of operation, SPIRIT would make these measurements more slowly than SAFIR. However, the exposure times would be quite reasonable ($\sim 10^5$ s) if SPIRIT, like SAFIR, uses ultra-sensitive detectors and cryogenic ($\sim 4$ K) optics, and if the SPIRIT collecting mirrors are about 3 m in diameter. The SPIRIT optical delay line should be sized to provide $R \sim 10^3$ spectroscopy over a several arcminute field of view. The wavelength range covered by SPIRIT should extend to the shortest wavelength continuously accessible to ALMA through the atmosphere, $\sim 800$ \micron .

\end{document}